\documentclass[10pt, a4paper]{article}

\usepackage{graphicx}
\usepackage{subcaption}
\usepackage{authblk}
\usepackage{amsmath}
\usepackage[english]{babel}
\usepackage{array}

\usepackage{cite}
\usepackage{siunitx}
\begin{document}


\title{Erd{\H{o}}s-R{\'e}nyi phase transition in the Axelrod model on complete graphs}

\author[1,2]{Sebasti\'an Pinto \thanks{spinto@df.uba.ar}}
\author[1,2]{Pablo Balenzuela}

\affil[1]{\small{Departamento de F\'isica, Facultad de Ciencias Exactas y Naturales, Universidad de Buenos Aires. Av.Cantilo s/n, Pabell\'on 1, Ciudad Universitaria, 1428, Buenos Aires, Argentina.}}
\affil[2]{\small{Instituto de F\'isica de Buenos Aires (IFIBA), CONICET. Av.Cantilo s/n, Pabell\'on 1, Ciudad Universitaria, 1428, Buenos Aires, Argentina.}}

\date{\today}
             
\maketitle

\begin{abstract}
\par The Axelrod model has been widely studied since its proposal for social influence and cultural dissemination. In particular, the community of  statistical physics focused on the presence of a phase transition as a function of its two main parameters, $F$ and $Q$. 
In this work, we show that the Axelrod model undergoes a second order phase transition in the limit of $F \rightarrow \infty $ on a complete graph. This transition is equivalent to the Erd{\H{o}}s-R{\'e}nyi phase transition in random networks when it is described in terms of the probability of interaction at the initial state, which depends on a scaling relation between $F$ and $Q$. We also found that this probability plays a key role in sparse topologies by collapsing the transition curves for different values of the parameter $F$.

\end{abstract}


\section{Introduction}

\par The Axelrod model, originally proposed for cultural dissemination \cite{axelrod1997dissemination}, is grounded in two key dynamical features: Social influence, through which people become more similar when they interact; and homophily, which is the tendency of individuals to interact preferentially with similar ones.
Specifically, the agents are described by  a vector of $F$ components called cultural features, which can take one of $Q$ integer values called cultural traits.  
The dynamics of the model is based on an imitation rule: A random agent adopts a cultural trait of another one with a probability proportional to the number of shared features. 
\par Despite its simplicity, the Axelrod model attracted the attention of the  statistical physics community due to the emergency of a phase transition from a monocultural  to a multicultural state \cite{castellano2000nonequilibrium, klemm2003nonequilibrium}.
The phase transition takes place by varying the number of cultural traits $Q$ for a given fixed $F$. If the number of cultural traits is low, the probability of interaction is high, leading the system to a monocultural state. If $Q$ is high, the mentioned probability is low and after few interactions the system evolves to a stationary multicultural state.
\par This phase transition is usually studied by taking the size of the biggest fragment as the order parameter. The transition was reported to be continuous for one-dimensional networks and discontinuous for two dimensions when $F > 2$ \cite{klemm2003role}, although a continuous transition is recovered when the topology becomes small-world \cite{klemm2003nonequilibrium}.
On the other hand, for $F = 2$ , the type of the transition is the opposite: Continuous for 2-D, and discontinuous for small-world networks \cite{reia2016effect}.
The case of $F = 2$ is important due to the possibility of taking an analytical approach to study the model \cite{vazquez2007non}.
\par Several scaling relationships have been found in  the Axelrod model. For instance, in \cite{peres2015nature} and \cite{reia2016phase} a finite-size scaling analysis is performed for $F=2$ in square-lattices and small-world networks. A scaling relation between $Q$ and size $N$ can be found in scale-free networks for $F=10$  \cite{klemm2003nonequilibrium}, a scaling relationship between the density of active bonds and time in one-dimensional network in \cite{vilone2002ordering}, and the finding of an effective noise rate is explored in \cite{klemm2003global}.
\par Among the reported scaling relations, there is a particular one reported in \cite{klemm2005globalization} and \cite{lanchier2013fixation} where the transition curves in one-dimensional networks collapse when the control parameter is  $F/Q$. This ratio has an immediate interpretation as the mean value of shared features given two agents in the initial state, suggesting that the initial distribution contains key information about the final outcome of the Axelrod model. 
\par In this work, we review the Axelrod model in terms of the initial interaction probability between agents ($p_{int}$).  In particular, we found that the second order phase transition in the limit of $F \rightarrow \infty $ on a complete graph is equivalent to the phase transition observed in Erd{\H{o}}s-R{\'e}nyi random networks \cite{erdHos1960evolution}. 
Notoriously, we also observed that this interaction probability between agents ($p_{int}$) plays a key role in describing the Axelrod model on sparse topologies, by making the transition curves collapse for different values of $F$. 

\section{Models}

\subsection{Axelrod model}

\par The Axelrod model \cite{axelrod1997dissemination} describes each agent by a vector of $F$ components, which represents a set of cultural features.
The initial state of the system is set by assigning with equal probability one of $Q$ integer values to each feature. 
The value of $Q$ represents the number of different cultural traits that a particular feature adopts.
Once the initial condition is set, the dynamics of the system is based on a pairwise interaction mechanism, which relies on two fundamental hypothesis:
\begin{itemize}
\item Homophily: The probability of interaction between two individuals is proportional to their cultural similarity, that is, the number of features they share.
\item Social Influence: After each interaction, the agents become more similar. It means that one of the agents copies a feature from the other which they previously did not share.
\end{itemize} 
This model shows a non-equilibrium phase transition from a monocultural to a multicultural state by varying the value of $Q$ for a fixed $F$. When $Q < Q_c$, the probability that two agents can interact since the initial state is high, so all agents end with the same cultural vector.
On the other hand, when $Q > Q_c$, the probability of interaction at the initial state is low so the final state shows a coexistence of regions with different cultural states.
\par The transition is usually characterized by measuring the size of the biggest fragment. We define a fragment as a group of topologically connected agents with at least one feature in common. At the final state, the biggest fragment is made up only by agents with the same cultural state.

\subsection{Erd{\H{o}}s-R{\'e}nyi graph}
\label{sec:erdos_renyi}

\par The Erd{\H{o}}s-R{\'e}nyi network is a graph model in which a set of $N$ initially disconnected nodes are linked with probability $p$. 
The outcome of this model is a network with a binomial degree distribution of parameters $N$ and $p$, with mean degree $ k = (N-1)p$.
\par An interesting feature of this model is that the size of the largest connected component shows a second-order phase transition at $ k_c = 1$: When $k < k_c$ the largest component has a finite size, while for $k > k_c$ it scales with the size of the system \cite{erdHos1960evolution, newman2003structure}. 

\subsection{Connection between both models}
\label{sec:prior}

\par Given two agents, the parameters $F$ and $Q$ set their initial number of shared features by sampling this quantity from a binomial distribution with parameters $F$ and $1/Q$. 
If we define the initial interaction probability $p_{int}$ as the probability that two agents are able to interact at the initial state (i.e, the probability of sharing at least one feature), then:
\begin{equation}
    p_{int} = 1 - (1 - \frac{1}{Q})^F
    \label{eq:pint}
\end{equation}
Given this definition, we can think the initial state of the Axelrod model on a complete graph as an Erd{\H{o}}s-R{\'e}nyi model with parameter $p_{int}$.
In other words, we discriminate between links with zero and non-zero homophily, being $p_{int}$ equal to the fraction of pairs of agents that can interact at the initial state.

\section{Results}

\subsection{Axelrod model in complete graphs}

\par
\begin{figure}[ht]
\centering
\includegraphics[width = \columnwidth]{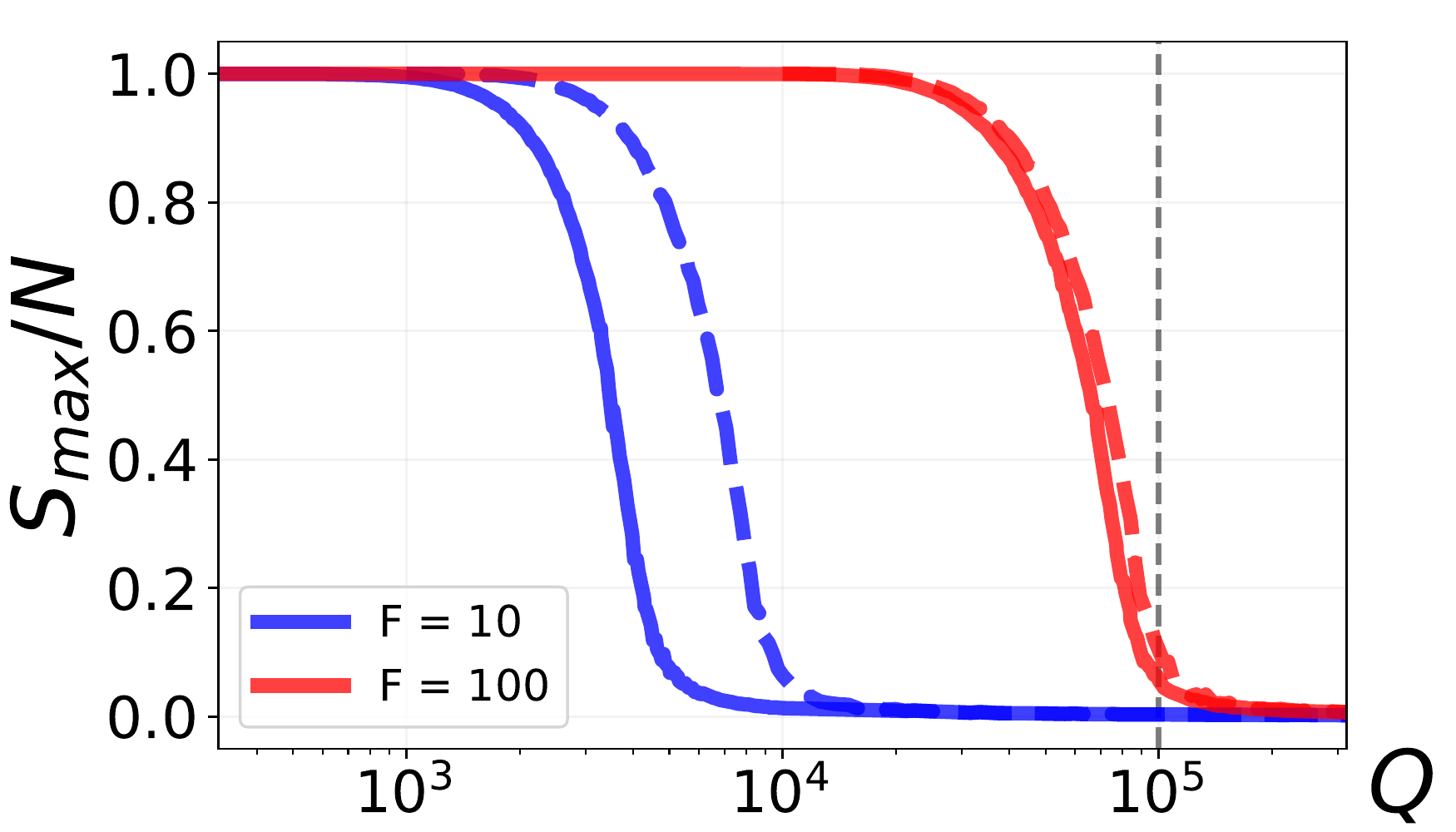}
\caption{{\bf Axelrod transition in a complete graph for different values of $F$.} Solid lines show the relative size of the biggest fragment $S_{max}/N$ as a function of the parameter $Q$ for $N = 1024$ at the final state. Dashed lines show their initial values for each $F$. When $F$ increases, $S_{max}/N$ set by the Axelrod dynamics becomes closer to its initial value. Vertical dashed lines points the critical value $Q_c=10^5$ mentioned in the text.
}
\label{figure1}
\end{figure}

 We first analyze the Axelrod model on a complete graph of $N=1024$ agents. Fig. \ref{figure1} shows that the relative size of the biggest fragment ($S_{max}/N$) in the final state is smaller than its initial value for low $F$. However, this difference approaches to zero when $F$ increases.
This result suggests that the relative size of the stationary biggest fragment is fully determined at the initial condition in the limit of $F \rightarrow \infty$.
The importance of the initial condition is reflected in the fact that two agents who  initially  do not share any feature cannot interact, at least until other interactions take place and eventually change their cultural states. 

\par
\begin{figure}[ht]
    \centering
    \includegraphics[width=\columnwidth]{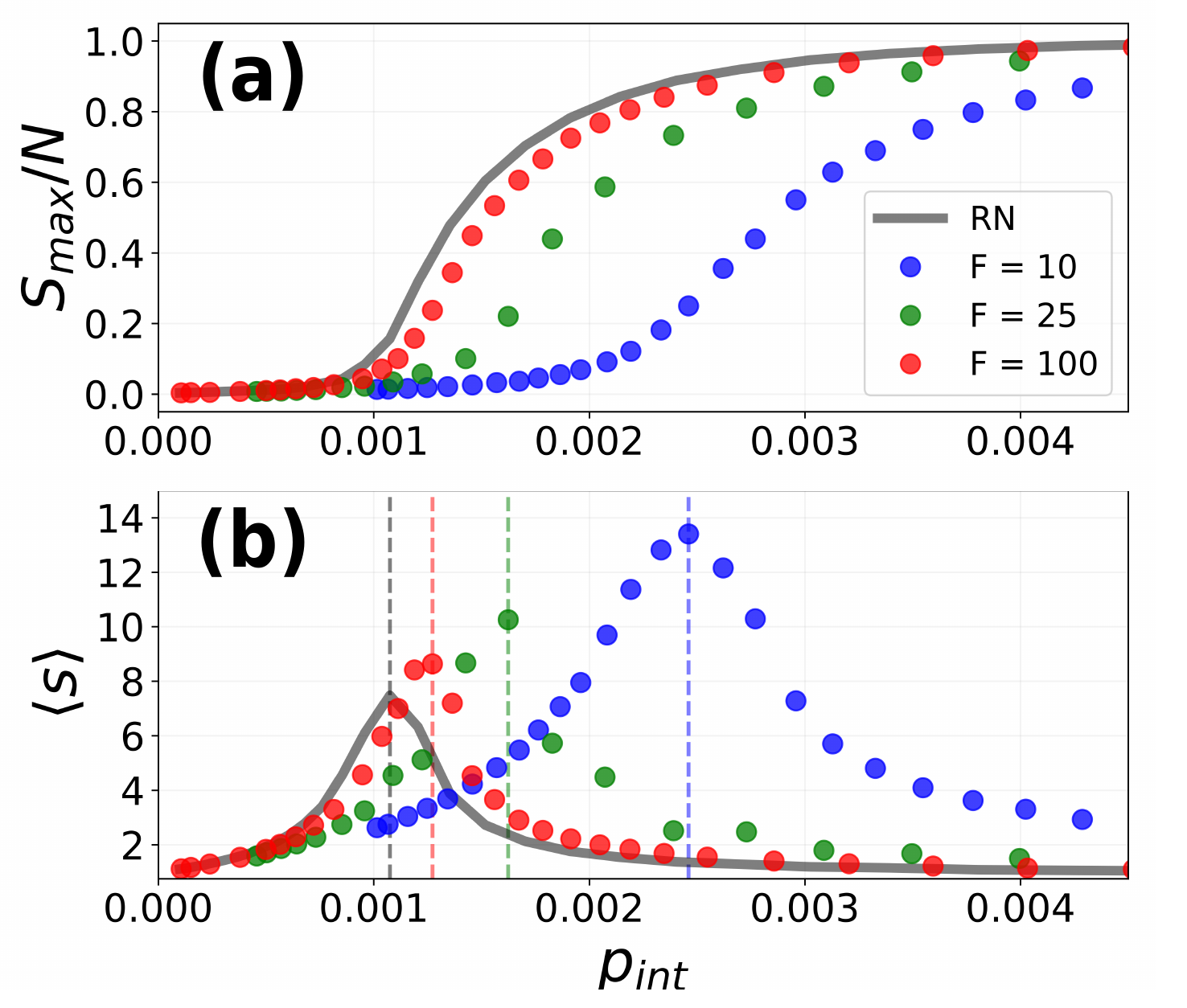}
    \caption{{\bf Relative size of biggest fragment $S_{max}/N$ and average fragment size $\langle s \rangle$ as function of $p_{int}$ for $N = 1024$.} Both figures show that the Axelrod transition tends to the Erd{\H{o}}s-R{\'e}nyi (RN) transition for increasing $F$.
    Dashed lines in panel (b) point out the critical values of $p_{int}^c(N)$ for different $F$ and the random network model.}
    \label{figure2}
\end{figure}

 The suggestion of the equivalence between both transitions in the limit of $F \rightarrow \infty$ is more clear by taking $p_{int}$ as the control parameter.
Fig. \ref{figure2} shows this for the relative size of the biggest fragment (panel (a)) and the average finite-fragment size (panel (b)).
Moreover, the definition of $p_{int}$ (Eq. (\ref{eq:pint})) allows to estimate the critical value of the transition $Q_c$ for large $F$. Since the biggest fragment emerges in an Erd{\H{o}}s-R{\'e}nyi network when $Np_{int}^c \sim 1$ \cite{newman2003structure},
\begin{equation*}
Q_c \sim (1 - (1 - \frac{1}{N})^\frac{1}{F})^{-1},
\end{equation*}
in the Axelrod model.  
Applying this analogy for $N = 1024$ and $F = 100$, it gives $Q_c\sim 10^5$, as can be seen in Fig. \ref{figure1}. 

\subsubsection{Critical point and exponents}

\par
\begin{figure}[ht]
    \centering
    \includegraphics[width=\columnwidth]{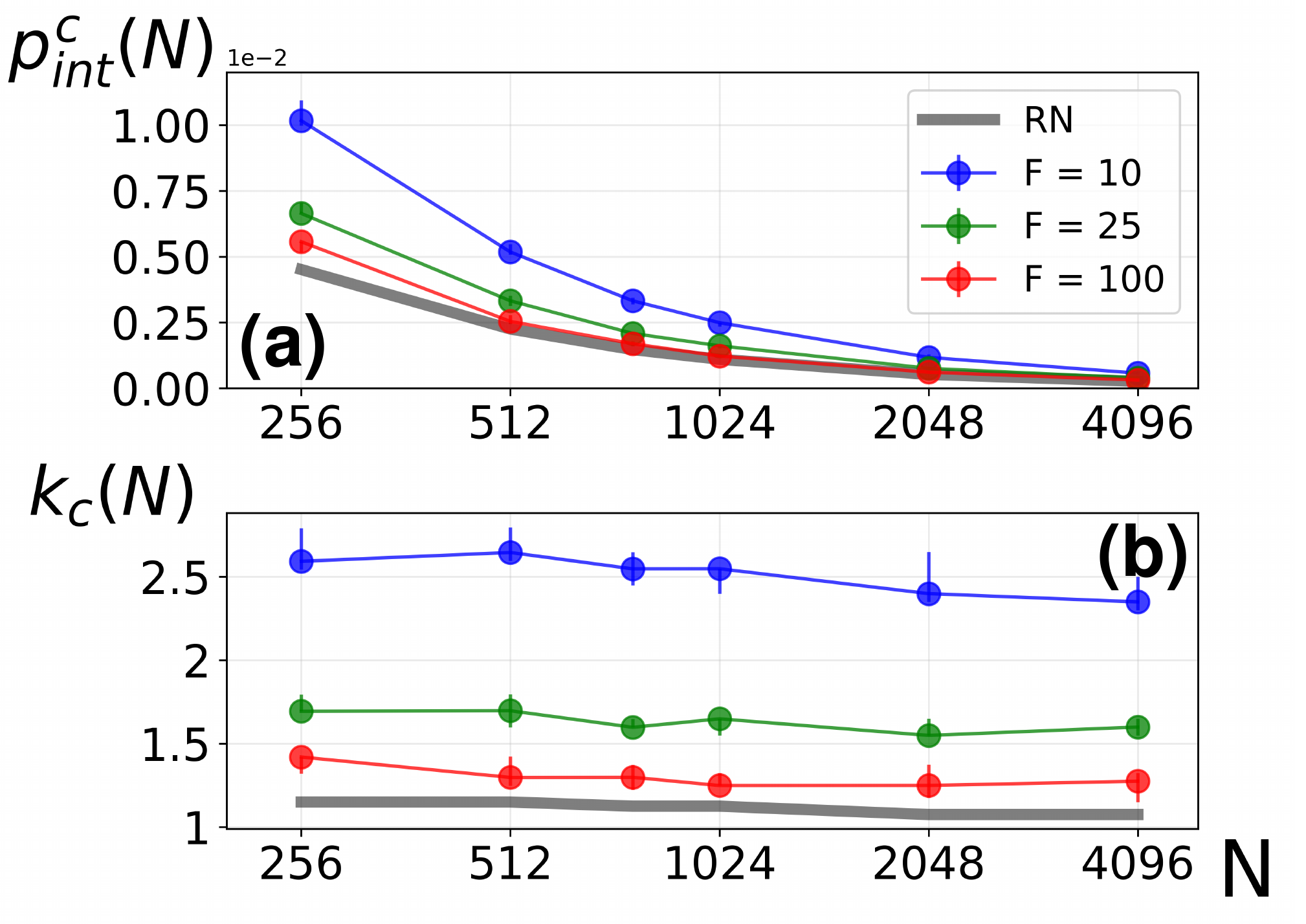}    
    \caption{{\bf Critical point and mean degree for different values of  $F$.} It can be observed that when $F$ is increased, $p_{int}^c(N)$ tends to the random network critical point at fixed $N$, while $k_c$ tends to $1$ for all $N$.}
    \label{figure3}
\end{figure}

 As can be observed in Fig. \ref{figure2}, the critical point tends to the Erd{\H{o}}s-R{\'e}nyi one for large $F$.
Fig. \ref{figure3} shows that $p_{int}^c(N)$ effectively tends to the corresponding value in random networks when $F$ is increased at fixed $N$ (panel (a)). 
Although the differences between $F$ values seems to vanish when $N \to \infty$, if we define the mean degree
\begin{equation*}
    k_c = (N-1)p_{int}^c(N),
\end{equation*}
as usual for the Erd{\H{o}}s-R{\'e}nyi model, we can see in panel (b)  that $k_c(F)$ is stable as a function of $N$ and tends to the theoretical value $k_c=1$ for large $F$ (see section \ref{sec:erdos_renyi}).
In the case of the Axelrod model, $k$ is interpreted as the average number of neighbours that a given agent can interact at the initial state.

\par The equivalence between Erd{\H{o}}s-R{\'e}nyi's and Axelrod model phase-transition's for large $F$ can be completed by the calculation of the critical point and critical exponents in the thermodynamics limit. 
Due to the closeness of the transition curves observed in Fig. \ref{figure2}, we estimate them for $F=100$ by applying finite-size scaling according to the procedure sketched in \cite{newman1999monte}. We provide details of these calculations in Appendix \ref{sec:Finite_size_scaling}.
\par Table \ref{tab:table1} shows that our estimations of the critical exponents are consistent with the theoretical values predicted for the Erd{\H{o}}s-R{\'e}nyi model \cite{newman2003structure}. 
We can observe that in all cases, the theoretical value is included in the $95\%$ confidence interval of our estimation.  Moreover, applying the finite-size scaling methodology to finite random networks within the same range of $N$, we found that the matching between both models is even stronger.
\par Finally, our estimation of the critical point was 
$p_{int}^c(N = \infty) =  7 \times 10^{-5}$, which is consistent with the theoretical value of $0$ within a $95 \%$ confidence interval (see Table \ref{tab:table1}).

\begin{table}[ht]
\centering
    \begin{tabular}{c|c|c|c|c}
         & \bf Theoretical & \bf Axelrod & \bf 95\% CI & \bf RN 95\% CI \\
         \hline
         \multicolumn{5}{l}{Critical exponents} \\
         \hline
         $\nu$ & 1 & 0.88 & 0.85 - 1.01 & 0.87 - 1.02\\
         $\beta/\nu$  & $1/3$ & 0.42 & 0.11 - 0.52 & 0.20 - 0.38\\
         $\gamma/\nu$ & $1/3$ & 0.36 & 0.33 - 0.46 & 0.33 - 0.35\\
         $\tau$ & $5/2$ & 2.45 & 2.41 - 2.61 & 2.49 - 2.53\\
         \hline
         \multicolumn{5}{l}{Critical point} \\
         \hline
         $p_{int}^c$ & $0$ & \num{7e-5} & (-7 - 10)\num{e-5} & (-5 - 9)\num{e-5}\\
         \hline
    \end{tabular}
    \caption{{\bf Critical exponents and critical point}. Theoretical values for the Erd{\H{o}}s-R{\'e}nyi phase transition and estimations from the Axelrod model with $F=100$, with their respective 95\% confidence intervals (CI). We also report the 95\% confidence intervals of the random network (RN) model applying finite-size scaling within the same range of $N$.}
    \label{tab:table1}
\end{table}

\subsubsection{Other topological features}

\par
\begin{figure}[ht]
    \centering
    \includegraphics[width = \columnwidth]{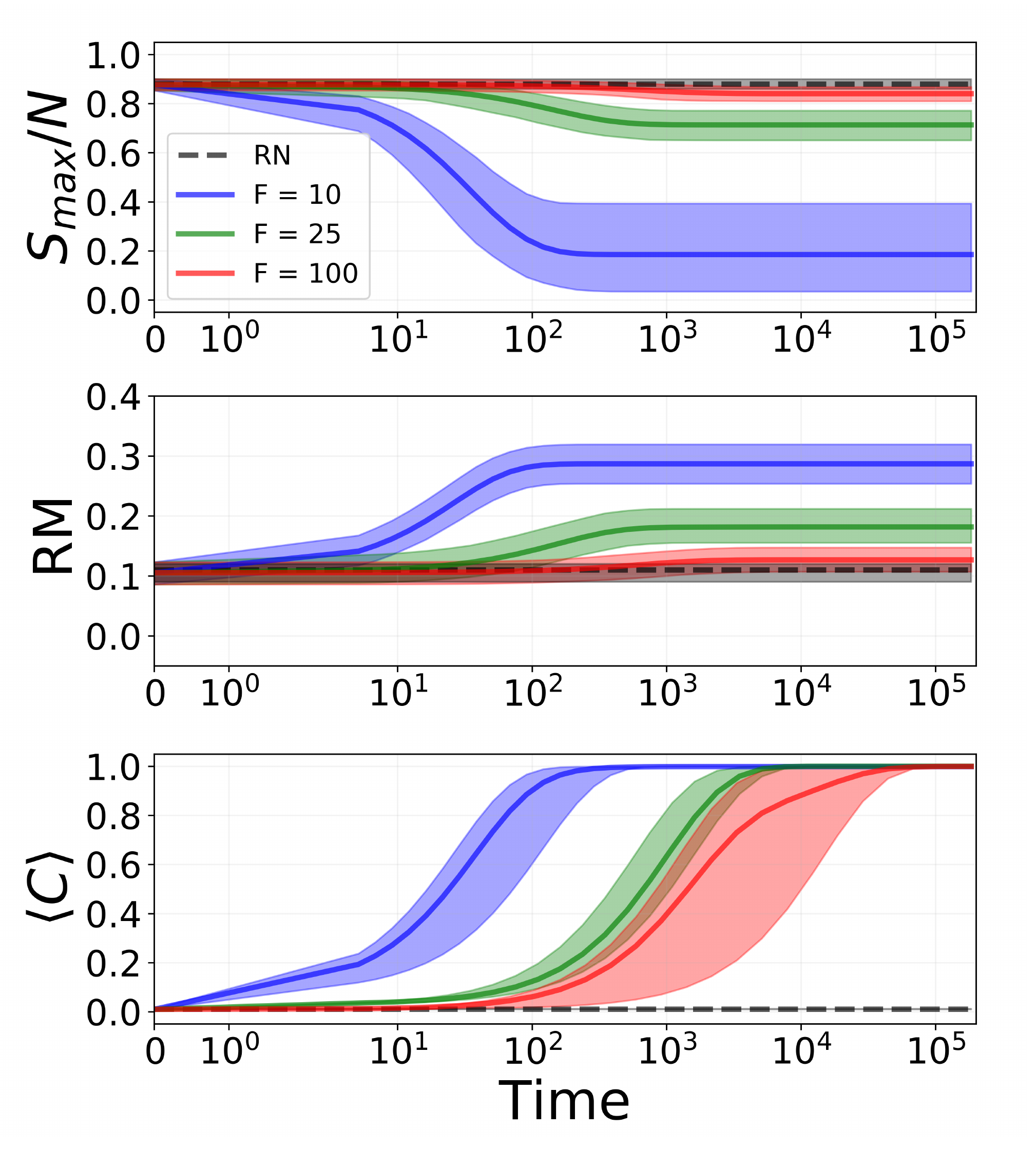}
    \caption{{\bf Relative size of the biggest $S_{max}/N$, relative multiplicity of fragments $RM$ and average clustering coefficient $\langle C \rangle$ of the biggest fragment}. This simulations belong to $N = 1024$ and $p_{int} = 0.0024$. Similar results were observed for different values of $p_{int}$. Dashed lines point out random network (RN) respective values. Time is measured per $N$ interactions.}
    \label{figure4}
\end{figure}

 It should be noticed that the equivalence between the Erd{\H{o}}s-R{\'e}nyi and the Axelrod model phase-transition for large $F$ (described by similar critical exponents) will not necessarily be present in other topological features given the dynamical evolution of the Axelrod model. 
Fig. \ref{figure4} shows the time evolution of $S_{max}/N$, the relative multiplicity $RM$, defined as the number of fragments normalized by $N$, and the average clustering coefficient of the biggest fragment $\langle C \rangle$, for different values of $F$  at fixed $p_{int}$. We also show the expected value of  the Erd{\H{o}}s-R{\'e}nyi model at the same probability.
It can be noticed that for large values of $F$, $S_{max}/N$ and $RM$ tend to be constant and closer to their initial value which is similar to the random graph model (pointed out by dashed lines in the figure). On the other hand, $\langle C \rangle$ always ends up reaching the value of $1$ (in this case, $\langle C \rangle \sim 0$ for Erd{\H{o}}s-R{\'e}nyi).
The observed behavior of $S_{max}/N$ and $RM$  tells us that the fragments at the initial state tend to break apart during the dynamics, while the increment of the clustering coefficient points that the connected components become cliques at the final state.
\par The evolution of these quantities could be understood as follows. First, we say that a link is active if it connects two agents with at least one feature in common (i.e a non-zero homophily link). Given an active link between two agents, the Axelrod model always tends to increase their similarity.
An active link can only become inactive  by third party interactions. When the value of  $F$ increases, the probability that an active link becomes inactive decreases and goes to zero when $F \rightarrow \infty$.
In this limit, the initial active links define the sizes of the connected components (as in Erd{\H{o}}s-R{\'e}nyi model) and the only effect of the dynamics is to transform all the connected components in cliques of the same size.

\subsection{Axelrod model in 2D lattices}

\par
\begin{figure}[ht]
\includegraphics[width = \columnwidth]{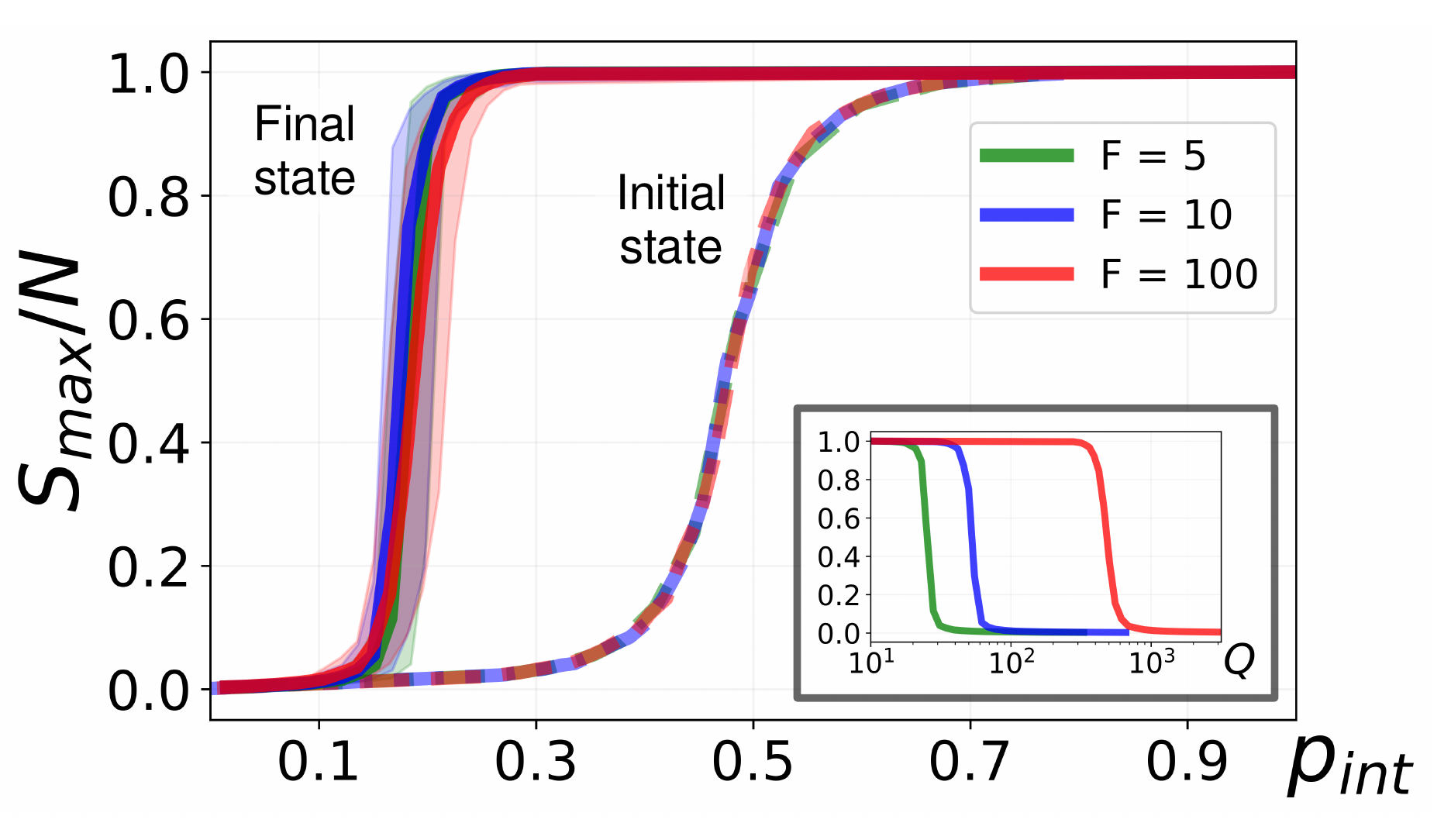}
\caption{{\bf Axelrod transition in a 2D lattice.} Relative size of the biggest fragment as a function of $p_{int}$ for both initial (dashes lines) and final state (solid lines with error bars). Same quantity as a function of $Q$ for different values of $F$ are shown in the inset. $N = 1024$.}
\label{figure5}
\end{figure}

 Let's explore now the Axelrod transition in a two-dimensional lattice as a function of the new control parameter $p_{int}$. Fig. \ref{figure5} shows the transition curves for different values of $F$ as a function of $p_{int}$, in addition to the initial relative size of the biggest fragment. The inset of this figure shows the same curves as function of $Q$, where it can be seen that the transition shifts to larger values of $Q$ when $F$ increases. 
\par Fig. \ref{figure5} shows that all transition curves collapse to one within the error bars. 
This collapse is also found in other sparse topologies as random regular networks with equivalent mean degree (not shown). 
In contrast to the observed behavior in complete networks (see Fig. \ref{figure1}), the collapsed curve does not match  the corresponding to the initial state. 
\par The collapse as a function of $p_{int}$ is essentially the same pointed out by \cite{klemm2005globalization} for one dimensional networks. As was mentioned in section \ref{sec:prior}, $F$ and $Q$ set the number of shared features for a pair of agents by sampling this quantity from a binomial distribution with parameters $F$ and $1/Q$. These quantities also set the value of $p_{int}$. In the limit of large $F$ and $Q$, this binomial distribution can be well approximated by a Poisson one with parameter $F/Q$, which is the mean shared features by two random agents and the control parameter introduced in \cite{klemm2005globalization} for one-dimensional networks. 

\par We can conclude here that, for sparse topologies, the Axelrod model can be described by a control parameter related to the similarities among agents at the initial state, in particular to the fraction of non-zero homophily links ($p_{int}$). This is  related to what we have observed for complete graphs when $F$ is large. The perspectives of these observations are sketched in the conclusions of our work.

\section{Conclusions}
\label{sec:Conclusions}

\par In this work, we show that the Axelrod model on complete graphs displays a phase transition equivalent to the one observed in the Erd{\H{o}}s-R{\'e}nyi model when the order parameter is plotted as a function of  $p_{int}$, which is the probability that two agents share at least one feature in the initial state. This happens in the limit of $F \rightarrow \infty$. This claim is supported by the calculation of critical exponents and the critical point following the approach of finite size scaling sketched in  \cite{newman1999monte}. 
\par This equivalence can be understood by observing that when $F$ increases, the size of the fragments as well as the amount of them are preserved during the dynamics. 
In other words, the probability that the fragments set in the initial state break apart goes to zero for large $F$.
However, the Axelrod dynamics is displayed by leading to a stationary state where the connected components become also cliques. 
In summary, it means that  in the limit $F \to \infty$, the initial fragments which are similar to an Erd{\H{o}}s-R{\'e}nyi realization with parameter $p_{int}(F,Q)$ are preserved but transformed in cliques.
An implication of this equivalence is that the final state of the Axelrod model is highly predictable from its initial state for large $F$. A possible future research line could be to explore if this conclusion is extensible to other dense topologies such as random regular networks with high mean degree.
\par On the other hand, when the Axelrod model is studied in terms of $p_{int}$ on sparse graphs, we found the collapse of transition curves for different values of $F$. These collapsed curves do not coincide with the initial state, as did in complete networks for $F \rightarrow \infty$.
However, this means that both the transition for complete graphs for large $F$ and sparse topologies can be described in terms of a quantity related to the similarity between agents at the initial state, in particular to the fraction of links with non-zero homophily $p_{int}$.
A scope of future works will be based on rewriting the Axelrod model in terms of the similarity between agents in order to analyze the explicit dependence of this model on quantities related to the initial similarity distribution.

\section{Acknowledgements}
\par We thank Luc\'ia Pedraza, Juan Pablo Pinasco, and Ignacio Sticco for bringing us a critical revision of the manuscript.

\bibliographystyle{unsrt}

\begin{thebibliography}{10}

\bibitem{axelrod1997dissemination}
Robert Axelrod.
\newblock The dissemination of culture: A model with local convergence and
  global polarization.
\newblock {\em Journal of conflict resolution}, 41(2):203--226, 1997.

\bibitem{castellano2000nonequilibrium}
Claudio Castellano, Matteo Marsili, and Alessandro Vespignani.
\newblock Nonequilibrium phase transition in a model for social influence.
\newblock {\em Physical Review Letters}, 85(16):3536, 2000.

\bibitem{klemm2003nonequilibrium}
Konstantin Klemm, V\'ictor~M Egu\'iluz, Ra\'ul Toral, and Maxi San~Miguel.
\newblock Nonequilibrium transitions in complex networks: A model of social
  interaction.
\newblock {\em Physical Review E}, 67(2):026120, 2003.

\bibitem{klemm2003role}
Konstantin Klemm, V\'ictor~M Egu\'iluz, Ra\'ul Toral, and Maxi San~Miguel.
\newblock Role of dimensionality in axelrod's model for the dissemination of
  culture.
\newblock {\em Physica A: Statistical Mechanics and its Applications},
  327(1-2):1--5, 2003.

\bibitem{reia2016effect}
Sandro~M Reia and Jos\'e~F Fontanari.
\newblock Effect of long-range interactions on the phase transition of
  axelrod's model.
\newblock {\em Physical Review E}, 94(5):052149, 2016.

\bibitem{vazquez2007non}
Federico V{\'a}zquez and Sidney Redner.
\newblock Non-monotonicity and divergent time scale in axelrod model dynamics.
\newblock {\em EPL (Europhysics Letters)}, 78(1):18002, 2007.

\bibitem{peres2015nature}
Lucas~R Peres and Jos{\'e}~F Fontanari.
\newblock The nature of the continuous non-equilibrium phase transition of
  axelrod's model.
\newblock {\em EPL (Europhysics Letters)}, 111(5):58001, 2015.

\bibitem{reia2016phase}
Sandro~M Reia and Jos{\'e}~F Fontanari.
\newblock The phase transition of axelrod’s model revisited.
\newblock {\em CoRR}, 2016.

\bibitem{vilone2002ordering}
Daniele Vilone, Alessandro Vespignani, and Claudio Castellano.
\newblock Ordering phase transition in the one-dimensional axelrod model.
\newblock {\em The European Physical Journal B-Condensed Matter and Complex
  Systems}, 30(3):399--406, 2002.

\bibitem{klemm2003global}
Konstantin Klemm, Victor~M Egu{\'\i}luz, Ra{\'u}l Toral, and Maxi San~Miguel.
\newblock Global culture: A noise-induced transition in finite systems.
\newblock {\em Physical Review E}, 67(4):045101, 2003.

\bibitem{klemm2005globalization}
Konstantin Klemm, V\'ictor~M Egu\'iluz, Raul Toral, and Maxi San~Miguel.
\newblock Globalization, polarization and cultural drift.
\newblock {\em Journal of Economic Dynamics and Control}, 29(1-2):321--334,
  2005.

\bibitem{lanchier2013fixation}
Nicolas Lanchier, Stylianos Scarlatos, et~al.
\newblock Fixation in the one-dimensional axelrod model.
\newblock {\em The Annals of Applied Probability}, 23(6):2538--2559, 2013.

\bibitem{erdHos1960evolution}
Paul Erd{\H{o}}s and Alfr{\'e}d R{\'e}nyi.
\newblock On the evolution of random graphs.
\newblock {\em Publ. Math. Inst. Hung. Acad. Sci}, 5(1):17--60, 1960.

\bibitem{newman2003structure}
Mark~EJ Newman.
\newblock The structure and function of complex networks.
\newblock {\em SIAM review}, 45(2):167--256, 2003.

\bibitem{newman1999monte}
M~Newman and G~Barkema.
\newblock {\em Monte carlo methods in statistical physics}.
\newblock Oxford University Press: New York, USA, 1999.

\bibitem{clauset2009power}
Aaron Clauset, Cosma~Rohilla Shalizi, and Mark~EJ Newman.
\newblock Power-law distributions in empirical data.
\newblock {\em SIAM review}, 51(4):661--703, 2009.

\bibitem{wasserman2013all}%
L. Wasserman.
\newblock All of statistics: a
  concise course in statistical inference.
\newblock Springer Science \& Business Media, 2013.

\end{thebibliography}

\appendix
\renewcommand\thefigure{\thesection.\arabic{figure}} 
\section{Critical exponents and finite size scaling}
\label{sec:Finite_size_scaling}
\setcounter{figure}{0} 

\par The critical exponents \cite{newman2003structure} are introduced following the usual relationships. Near the critical point:
\begin{align*}
\frac{S_{max}}{N} \sim (p_{int} - p_{int}^c)^\beta, \\ 
\langle s \rangle \sim |p_{int} - p_{int}^c|^{-\gamma},           
\end{align*}
where $p_{int}^c$ is the critical probability, $S_{max}/N$ is the relative size of the biggest fragment and the order parameter, whose fluctuations are measured by $\langle s \rangle$ which is the average finite-fragment size.
\par For finite systems, the critical exponents can be calculated by performing  finite-size scaling following \cite{newman1999monte}. Here, the authors propose the following scaling relationships:
\begin{align*}
\frac{S_{max}}{N} = N^{-\frac{\beta}{\nu}} F_1[(p_{int}-p_{int}^c) N^{\frac{1}{\nu}}], \\
\langle s \rangle = N^{\frac{\gamma}{\nu}} F_2[(p_{int}-p_{int}^c) N^{\frac{1}{\nu}}],
\end{align*}
where $F_1$ and $F_2$ are unknown scaling functions, but with the property that $F_{1(2)}(x) \rightarrow constant$ when $x \rightarrow 0$ (which means, near the critical value). 
This implies that exactly at the critical point:
\begin{align}
    \frac{S_{max}}{N} \sim N^{-\frac{\beta}{\nu}}
    \label{eq:finite_size1},\\
    \langle s \rangle \sim N^{\frac{\gamma}{\nu}}
    \label{eq:finite_size2}
\end{align}
These expressions can be used to estimate the relation between exponents without knowledge about $F_1$ and $F_2$. 
On the other hand, the argument of these scaling functions defines another relationship  followed by the exponent $\nu$ and the critical value $p_{int}^c$, which can be read in the following equation:
\begin{equation}
    p_{int}^c(N) = p_{int}^c - b N^{-\frac{1}{\nu}},
    \label{eq:pc_scaling}
\end{equation}
where $p_{int}^c(N)$ is the pseudo-critical point in which $\langle s \rangle$ takes its maximum value, being $p_{int}^c = p_{int}^c(\infty)$.
Finally, the fragments size distribution $f(s)$ near the critical point follows a power-law distribution with parameter $\tau$ (Eq. (\ref{eq:tau})). This relation defines this last critical exponent, which was calculated following the methodology sketched in \cite{clauset2009power}.
\begin{equation}
    f(s) \sim s^{-\tau}
    \label{eq:tau}
\end{equation}

\par
\begin{figure}[ht]
    \centering
    \includegraphics[width=\columnwidth]{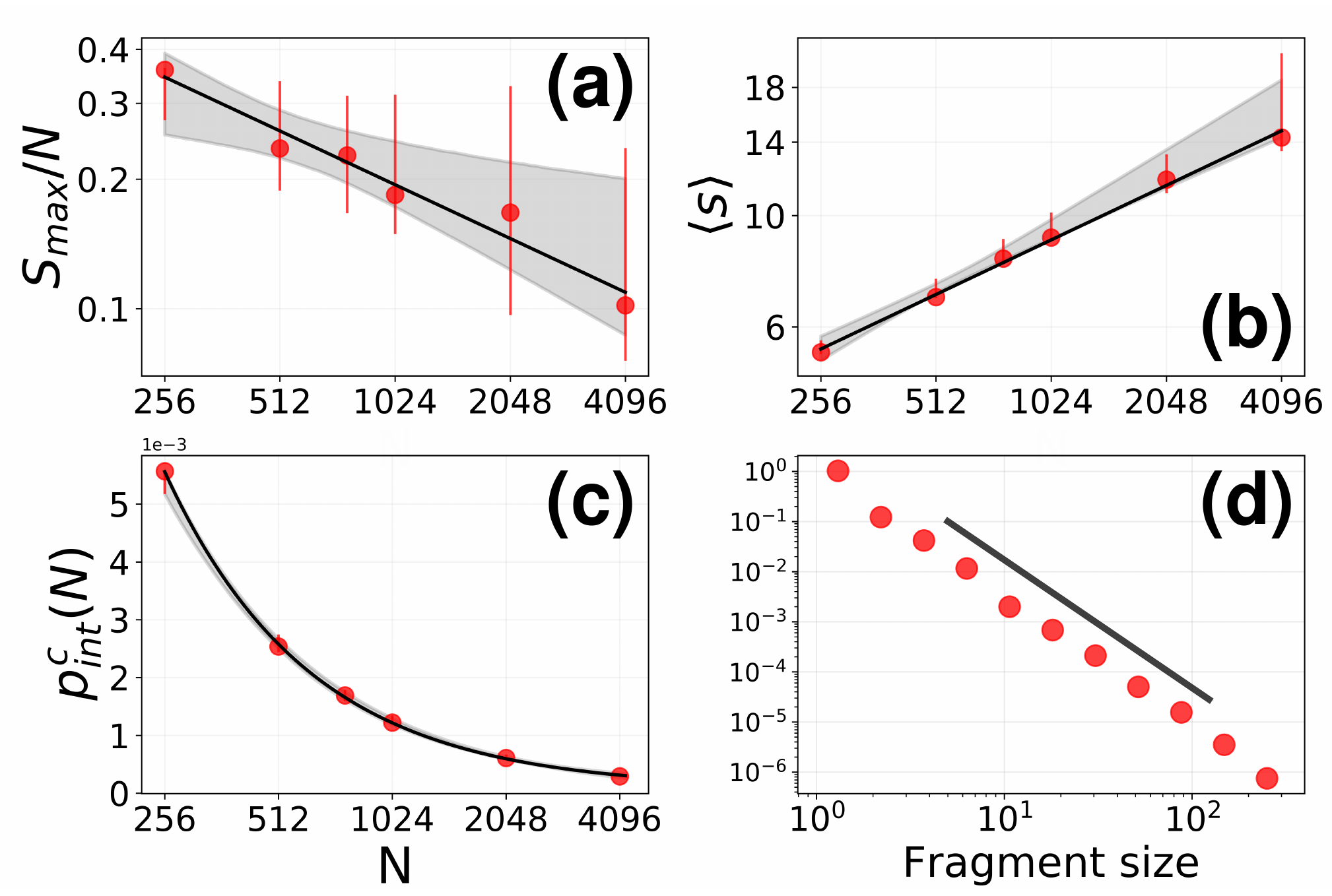}    
    \caption{{\bf Finite-size scaling and critical behaviour for $F=100$.} Red marks are point-estimates while full lines point out the fitted curves.
    Error bars and grey bands denote the 95-confidence interval of each estimation.
    Panel (d) belongs to the fragment distribution at the critical point with $N=4096$.}
    \label{fig:A1}
\end{figure}

 Figure \ref{fig:A1} shows an example of the calculus of the exponents for $F = 100$, which were reported in Table \ref{tab:table1}.
The relations of equations \ref{eq:finite_size1} to \ref{eq:tau} are respectively plotted in panels (a) to (d).
We also estimated the 95-confidence intervals of all quantities by bootstrapping \cite{wasserman2013all}, i.e. by recalculation them several times over different datasets generated by sampling with replacement the original measures taken from our simulations.
In the estimation of $p_{int}^c(\infty)$, we let this parameter to fluctuate around zero including negative values despite not being conceptually correct, in order to include zero in the confidence interval in the case it is supported by the statistics.
The same methodology was respectively implemented for each value of $F$ for the calculus of $p_{int}^c(N)$ of Fig. \ref{figure3}.

\end{document}